\documentclass[amsmath, amssymb, reprint, aps]{revtex4-1}

\usepackage{graphicx}
\usepackage{dcolumn}
\usepackage{bm}
\usepackage{color}

\newcommand{\lessgtrsim}[1]{\raisebox{0.05em}{\parbox[][][t]{0.7em}{\scalebox{0.8}[0.7]{$#1$}}}}

\newcommand{\sqrtHz}[0]{$\sqrt{\textrm{Hz}}$ }

\begin{document}


\title{All-Optical Vector Atomic Magnetometer}

 \author{B.\ Patton}
 \email{bpatton@berkeley.edu}
    \address{Department of Physics, University of California, Berkeley, CA 94720-7300}
    \address{Physik-Department, Technische Universit\"{a}t M\"{u}nchen, 85748 Garching, Germany}

 \author{E.\ Zhivun}
    \address{Department of Physics, University of California, Berkeley, CA 94720-7300}

 \author{D.\ C.\ Hovde}
    \address{Southwest Sciences Ohio Operations, Cincinnati, OH 45244}

 \author{D.\ Budker}
    \address{Department of Physics, University of California, Berkeley, CA 94720-7300}
    \address{Nuclear Science Division, Lawrence Berkeley National Laboratory, Berkeley, CA 94720}
    \address{Helmholtz Institute, Johannes Gutenberg University, 55099 Mainz, Germany}
\date{\today}

\begin{abstract}

We demonstrate an all-optical magnetometer capable of measuring the magnitude and direction of a magnetic field using nonlinear magneto-optical rotation in a cesium vapor.  Vector capability is added by effective modulation of the field along orthogonal axes and subsequent demodulation of the magnetic-resonance frequency.  This modulation is provided by the AC Stark shift induced by circularly polarized laser beams.  The sensor exhibits a demonstrated rms noise floor of 50 fT/\sqrtHz  in measurement of the field magnitude and 0.5 mrad/\sqrtHz in the field direction; elimination of technical noise would improve these sensitivities to 12 fT/\sqrtHz and 5 $\mu$rad/\sqrtHz\!\!, respectively.  Applications for a precise all-optical vector magnetometer would include magnetically sensitive fundamental physics experiments, such as the search for a permanent electric dipole moment of the neutron.

\end{abstract}

\maketitle

Spin-precession magnetometers \cite{OpticalMagnetometryBook, Budker_NP07} have found widespread application in disciplines ranging from geophysics \cite{Dang_APL10} to medicine \cite{Bison_APL09, Johnson_PMB13} and fundamental physics \cite{Vasilakis_PRL09, Altarev_PRL09}.  Alkali-vapor magnetometers in particular have experienced great advances in recent years, with sensitivities at or below the fT/\sqrtHz level demonstrated in the laboratory \cite{Dang_APL10, Ledbetter_PRA08, Griffith_OE10, Smullin_PRA09}.  Because these devices measure the Larmor precession frequency of atomic spins, they are intrinsically sensitive to the magnitude of an applied field rather than its projection along a particular direction.  This can be advantageous in that precision of the scalar field measurement is not limited by physical alignment of the sensors, as it can be in the case of triaxial fluxgates or superconducting quantum interference devices (SQUIDs).  Nevertheless, in many situations it is desirable to have full knowledge of a field's vector components.

There are several ways to derive vector field information from a scalar magnetometer.  In bias-field nulling, calibrated magnetic fields are imposed upon the magnetometer in order to achieve a zero-field magnetic resonance condition \cite{Slocum_IEEE63,Seltzer_APL04, Dong_IEEE13}.  With finite-field sensors using radiofrequency coils to drive the resonance (e.g., $M_x$ magnetometers \cite{Bloom_AO62}), one may add secondary continuous light beams and measure their modulation to extract vector information \cite{Fairweather_JPE72,Vershovskii_TPL11}.  It is also possible to detect magnetically sensitive resonances in electromagnetically induced transparency (EIT) schemes; the amplitudes of different EIT resonances can yield information about the relative angle between the laser polarization and the field \cite{Lee_PRA98, Cox_PRA11}.  Synchronously pumped magnetometers employing atomic alignment can also yield partial vector information when the magnetic field is not wholly perpendicular to the linear polarization of the pump beam \cite{Pustelny_PRA06}.

Perhaps the simplest way to adapt a scalar magnetometer for vector measurements is to operate it in the finite-field regime (e.g., through synchronous optical pumping \cite{Bell_PRL61,Gawlik_APL06,Higbie_RSI06}) and apply time-varying fields to it.  By applying orthogonal fields modulated at different frequencies, it is possible to demodulate the magnetic-resonance signal and determine which applied fields add linearly with the ambient field and which add in quadrature with it \cite{Rasson_GT91,Gravrand_EPS01,Vershovskii_TP06}.  Although this is effective, there are some situations where this approach is infeasible or undesirable.  One example would be the case of remote magnetometry \cite{Patton_APL12,Higbie_PNAS11}, where it would be impractical to apply fields to a distant atomic sample.  A different limitation appears in certain precision physics applications, such as the search for a neutron electric dipole moment (nEDM) \cite{Baker_PRL06,Altarev_PRL09,Knowles_NIMPRA09, Altarev_NC12}.  In such experiments alkali-vapor magnetometers can reduce systematic error by providing crucial magnetic-field information, but only if these sensors do not themselves produce field contamination.  All-optical alkali-vapor magnetometers are particularly well suited for nEDM tests as they can be designed to produce no significant static or radiofrequency fields \cite{Patton_EDMmag}.

Here we demonstrate an all-optical vector magnetic sensor based upon nonlinear magneto-optical rotation in a cesium vapor.  The effective magnetic field seen by the atoms is modulated by AC Stark shifts (``light shifts'') induced by orthogonally propagating laser beams.  Since the light shift of a circularly polarized beam is analogous \footnote{See the Supplemental Information for a discussion of the difference between the shifts induced by a light-shift beam and a magnetic field.} to an effective magnetic field oriented along its propagation direction \cite{Mathur_PR68, Cohen-Tannoudji_PRA72, Moriyasu_PRL09}, a comparison of the Larmor frequency shifts induced by these beams yields a measurement of the field angle.  If technical noise were eliminated, this magnetometer would have 12 fT/\sqrtHz precision in measurement of the field magnitude and 5 $\mu$rad/\sqrtHz in the field direction.  

\begin{figure}[t]
\includegraphics[scale=0.75]{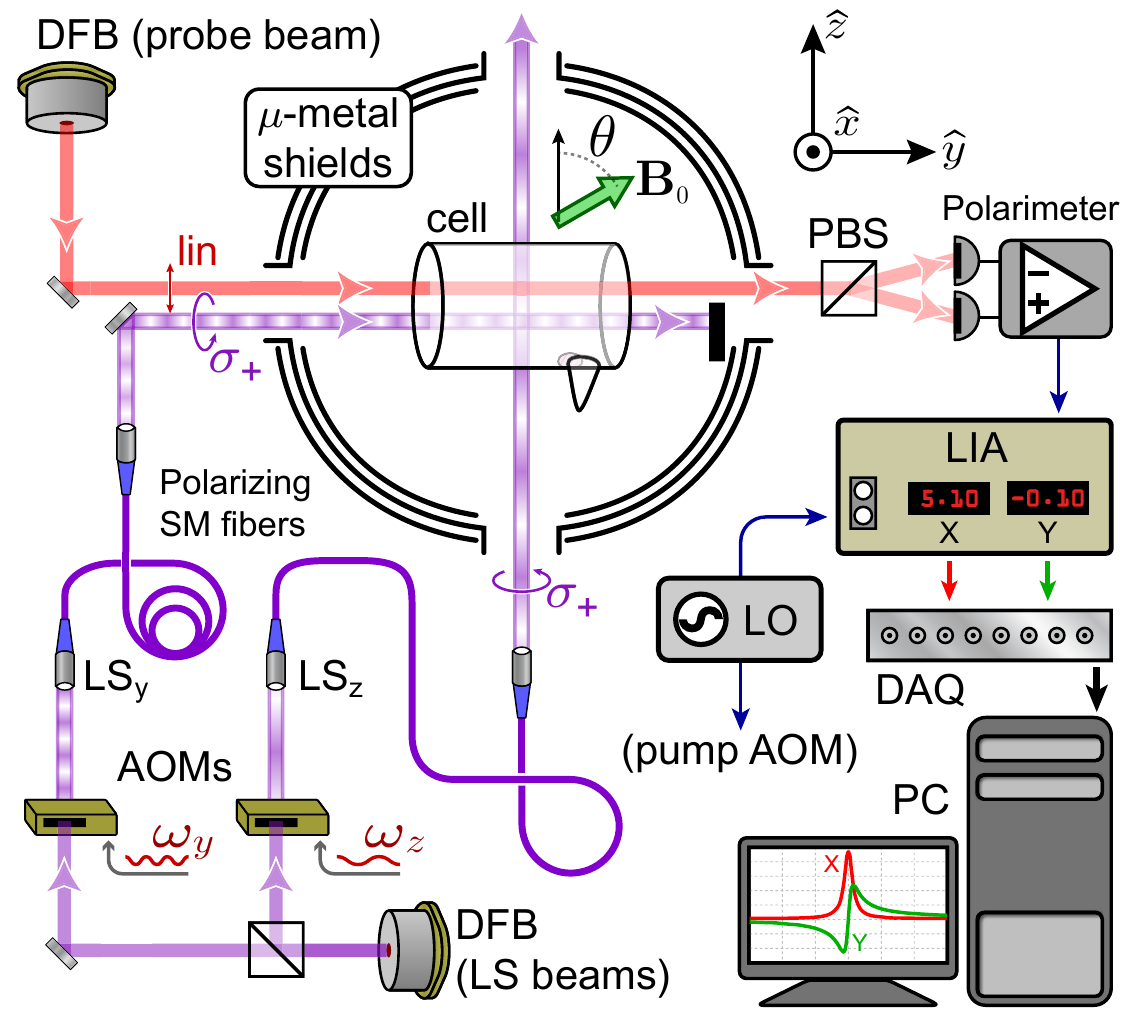}
{\caption[Vector magnetometry setup.]{\label{VectorMagSetup} Experimental schematic.  An amplitude-modulated, circularly polarized pump beam (not shown) propagates in the $\hat{x}$ direction.  The local oscillator (LO) controls the pump AOM and serves as a reference to the lock-in amplifier (LIA), whose analog output is recorded by a data acquisition card (DAQ) and read into a computer (PC).  A linearly polarized probe beam passes through the cell and is split by the polarizing beamsplitter (PBS) of a balanced polarimeter; the output of this polarimeter is demodulated by the lock-in.  Two circularly polarized light-shift beams $LS_y$ and $LS_z$ are independently modulated and sent through the cell along $\hat{y}$ and $\hat{z}$.  Coils allow the magnetic field $\textbf{B}_0$ to be tilted in the $\hat{y}$--$\hat{z}$ plane.}}
\end{figure}

The experimental setup is shown in Fig.~\ref{VectorMagSetup}.  The heart of the sensor is a cylindrical antirelaxation-coated \cite{Seltzer_JCP10} Cs vapor cell, approximately 5 cm diameter and 5 cm in length, with a longitudinal spin relaxation time of 0.7 seconds.  This cell is enclosed within four layers of $\mu$-metal magnetic shielding; measurements were performed at ambient temperature.  Coils wound on a frame within the innermost shield allow magnetic fields and gradients to be applied to the cell.  The field component oriented along $\hat{z}$ is produced by a current generated by a custom supply which can provide up to 150 mA (Magnicon GmbH).  This supply is housed in a temperature-stabilized enclosure and exhibits a relative drift of $\sim$10$^{-7}$ over 100 seconds.  A second current supply (Krohn Hite 523) is connected to the coil in the $\hat{y}$ direction, allowing the net field $\textbf{B}_0$ to be tilted in the $\hat{y}$--$\hat{z}$ plane.  The pump beam which drives the magnetic resonance is generated by a distributed feedback (DFB) diode laser that is locked with a dichroic atomic vapor laser lock (DAVLL) \cite{Yashchuk_RSI00} to the Cs $D1$ transition at 894 nm.  The $\hat{x}$-directed pump is circularly polarized and amplitude modulated with an acousto-optic modulator (AOM) at the $^{133}$Cs Larmor frequency $\omega_{L}$ to achieve synchronous optical pumping; the modulation waveform is a square wave with a duty cycle of 5\%.  A separate linearly polarized probe beam, generated by a DFB locked with a DAVLL to the Cs $D2$ transition, traverses the cell in the $\hat{y}$ direction.  The probe experiences optical rotation \cite{Budker_RMP02} in the polarized Cs sample, modulating its polarization at $\omega_{L}$.  This is detected by a balanced polarimeter with a differential transimpedance amplifier; its output is fed into a digital lock-in amplifier (Stanford Research Systems SR830) whose reference frequency is provided by the local oscillator which drives the pump AOM.  The phase of the lock-in amplifier is chosen such that the \textsf{X}(\textsf{Y}) output displays an absorptive (dispersive) Lorentzian as the driving frequency is scanned across the resonance.  Directly on resonance, the \textsf{X} output is maximum and the \textsf{Y} output is nulled; small shifts in the magnetic-resonance frequency $\omega_L$ cause a linear change in the \textsf{Y} output about zero.  With a time-averaged pump power of 2.5 $\mu$W and a probe power of 10 $\mu$W, the peak optical rotation signal is 5 mrad and the magnetic-resonance linewidth is 2.9 Hz.  The dominant contributions to this linewidth are alkali--alkali spin-exchange broadening and slight power broadening due to the pump and probe beams.  The beam powers and optical detunings were chosen to optimize the scalar sensitivity of the magnetometer.

In addition to the pump and probe, a third DFB laser tuned near the Cs $D2$ transition can be used to apply light-shift beams $LS_y$ and $LS_z$ in the $\hat{y}$ and $\hat{z}$ directions.  The optical frequency of the light-shift laser is actively controlled using a wavelength meter (\AA ngstrom/HighFinesse WS-7) and computer control of the laser current.  An optimal detuning of $\sim$5 GHz blue-shifted from the center of the $F\!\!=\!4 \rightarrow F'\!\!=\!5$ $D2$ transition was chosen to allow a large effective magnetic field ($\sim$1 nT/mW) with minimal ($\lessgtrsim{\lesssim}$0.5 Hz/mW) broadening of the magnetic-resonance line.  This beam is split into two paths and sent through independent AOMs, then coupled into two polarizing \footnote{Unlike conventional polarization-maintaining fiber, the HB830Z only transmits light which is linearly polarized along one of the axes of the anisotropic fiber core; the other polarization experiences large attenuation.} fiber patch cables (Fibercore HB830Z).  After the fibers, the light-shift beams are sent through quarter-wave plates to generate circularly polarized beams which pass through the cell along the $\hat{y}$ and $\hat{z}$ axes.  Optical pickoffs (not shown in Fig.~\ref{VectorMagSetup}) and photodiodes directly before the shields allow the power of each light-shift beam to be measured.  In an evacuated antirelaxation-coated cell, the alkali atoms rapidly sample the internal volume of the cell and experience a light shift equivalent to the volume-averaged intensity of the laser beam within the cell.  Thus two beams of the same power will possess slightly different light-shift coefficients (measured in nT/mW) when propagating in different directions due to asymmetry of the cell dimensions.  Nevertheless, their ratio will remain independent of the optical detuning of the light-shift laser.


\begin{figure}[t]
\includegraphics[scale=0.8]{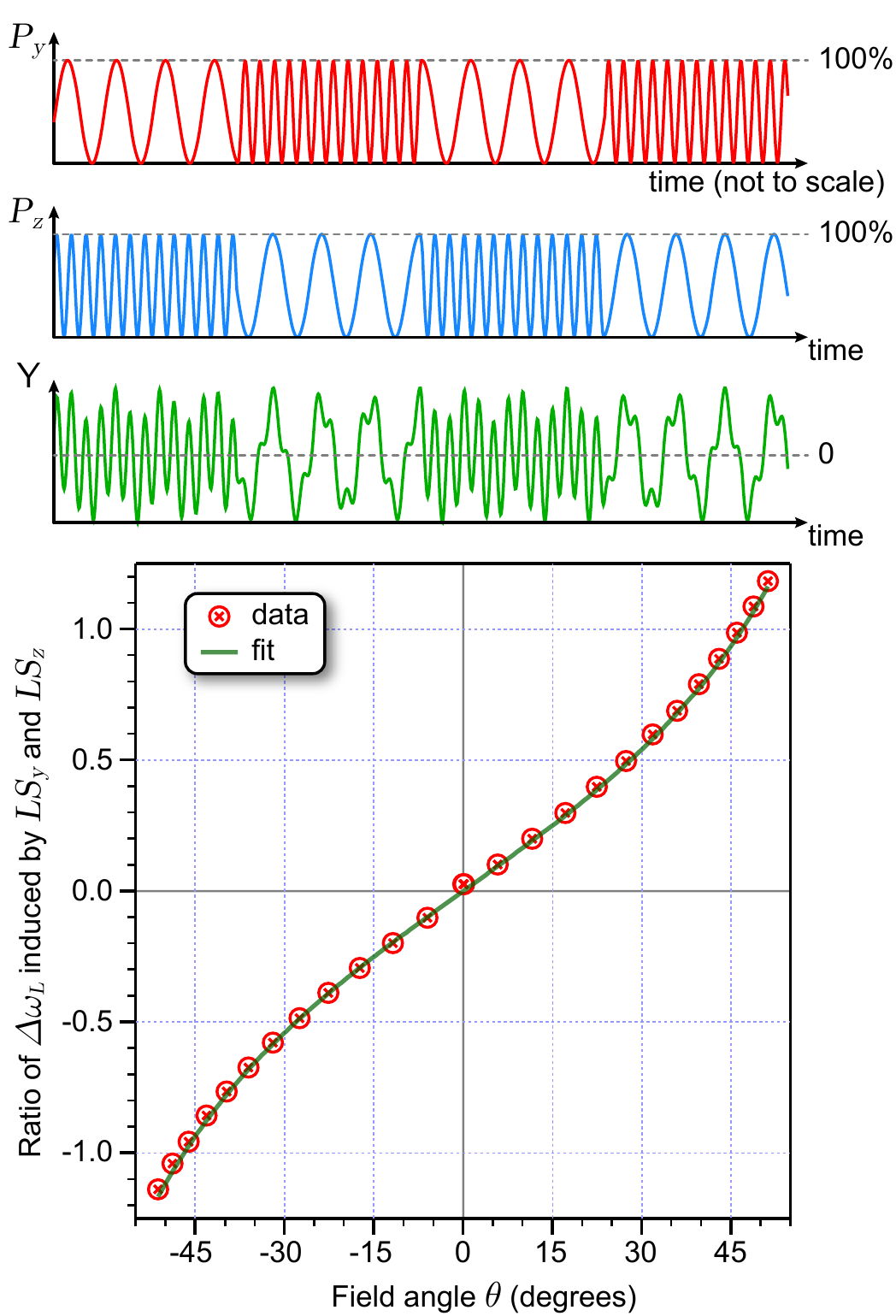}
{\caption[Vector magnetometry calibration.]{\label{tanplot} Above: Depiction of the $LS_y$ and $LS_z$ beam powers  versus time and the resulting (simulated) change in the lock-in \textsf{Y} output about zero.  Demodulation of the latter yields the contributions of  $LS_y$ and $LS_z$ to the shift in the magnetic-resonance frequency.  Below: The ratio of the Larmor frequency shift induced by the $LS_y$ and $LS_z$ beams, plotted as a function of field angle $\theta$ from the $\hat{z}$ axis.  The curve shows a fit to Eq.~(\ref{taneq}).  Each data point resulted from 20 seconds of averaging; uncertainties in the data points are uniformly below $10^{-2}$.}}
\end{figure}


To demonstrate the effective magnetic fields produced by $LS_y$ and $LS_z$, we recorded the data shown in Fig.~\ref{tanplot}.  For this measurement, the primary $\hat{z}$ field was held constant at 946.5 nT and an additional $\hat{y}$ field was varied between -1180.5 nT and +1177 nT.  Thus the field's magnitude $B_0$ changed with its angle $\theta$ in the $\hat{y}$--$\hat{z}$ plane, requiring the local oscillator and the lock-in reference phase to be reset for each measurement.  At each field, the respective light shifts produced by the $LS_y$ and $LS_z$ beams were measured by modulating the two beam intensities at different frequencies (12 and 20 Hz) and demodulating the lock-in $\textsf{Y}$ output in software.  The average intensity of each light-shift beam was 0.5 mW.  Although it improves precision to measure the system response to both beams simultaneously, it is important to alternate the fast and slow modulation in each channel, as shown in Fig.~\ref{tanplot}.   This is because the atomic system acts as a low-pass filter for fast field perturbations, since it is in effect a driven oscillator with a damping rate on the order of the magnetic-resonance linewidth.  

Assume that the magnetometer is operated in the finite-field regime, such that the magnetic resonance frequency is much higher than the resonance linewidth.  The modulated $LS_y$ beam produces an effective magnetic field of magnitude $B_y = P_y \alpha_y [$\mbox{$^1\!/\!_2$} + \mbox{$^1\!/\!_2$} $\sin(\omega_y t)]$, where $P_y$ is the beam power, $\alpha_y$ is its effective light-shift coefficient, and $\omega_y$ the amplitude-modulation frequency. Similarly, $LS_z$ produces $B_z = P_z \alpha_z [$\mbox{$^1\!/\!_2$} + \mbox{$^1\!/\!_2$} $\sin(\omega_z t)]$.  To maintain the synchronous pumping condition, the fields $B_y$ and $B_z$ are assumed to be  comparable to the resonance linewidth (in field units).  Adding these fields to the vector components of $\textbf{B}_0$, the total field magnitude becomes:
%
\begin{eqnarray}
B_{\textrm{tot}} & = & B_0 \sqrt{1+2\frac{B_y \sin \theta + B_z \cos \theta}{B_0} + \frac{B_y^2 + B_z^2}{B_0^2}} \nonumber \\ 
& \approx & B_0 + B_y \sin \theta + B_z \cos \theta + \zeta,
\label{LSfield01}
\end{eqnarray}
where the approximation is valid for $B_y, B_z \ll B_0$ and the small quadratic correction $\zeta$ is given by:
\begin{equation}
\zeta = \frac{\left(B_y \cos \theta - B_z \sin \theta \right)^2}{2B_0}.
\label{zeta}
\end{equation}
Since the lock-in \textsf{Y} output is proportional to the change in effective Larmor frequency induced by the light-shift fields, demodulation of the signal at frequencies $\omega_y$ and $\omega_z$ will extract the terms in Eq.\ (\ref{LSfield01}) proportional to $B_y$ and $B_z$.  Thus the ratio of the measured light shifts is:
\begin{equation}
\frac{(\Delta B_{\textrm{tot}})_{LSy}}{(\Delta B_{\textrm{tot}})_{LSz}}  \approx  \frac{P_y \alpha_y}{P_z \alpha_z} \tan \theta.
\label{taneq}
\end{equation}
Here we have ignored the contribution from the terms in $\zeta$ and other terms of higher power in $(B_{y,z}/B_0)$, which cause modulation of $B_{\textrm{tot}}$ at harmonics other than $\omega_y$ and $\omega_z$ or scale by powers of $\left| B_{y,z}/B_0 \right|$ (here $\lessgtrsim{\lesssim}~10^{-3}$).

\begin{figure}[t]
\includegraphics[scale=0.8]{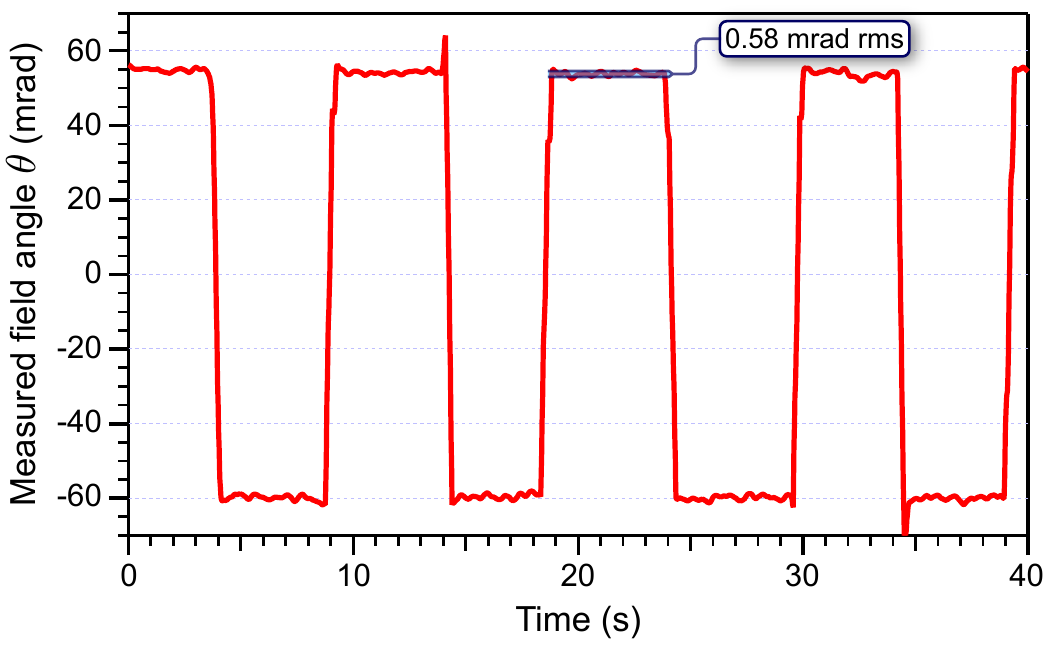}
{\caption[Precision of the vector measurement.]{\label{StepPlot} Measured field angle $\theta$ as a function of time while the applied $\hat{y}$ field is being switched.  The average rms noise for a constant field angle translates to a precision of 0.47 mrad/\sqrtHz in measurement of the field direction.  The steps in the plotted ratio are slightly low-pass filtered due to the time constants of the lock-in amplifier and the secondary demodulation at $\omega_y$ and $\omega_z$.}}
\end{figure}

\begin{figure}[t]
\includegraphics[scale=0.8]{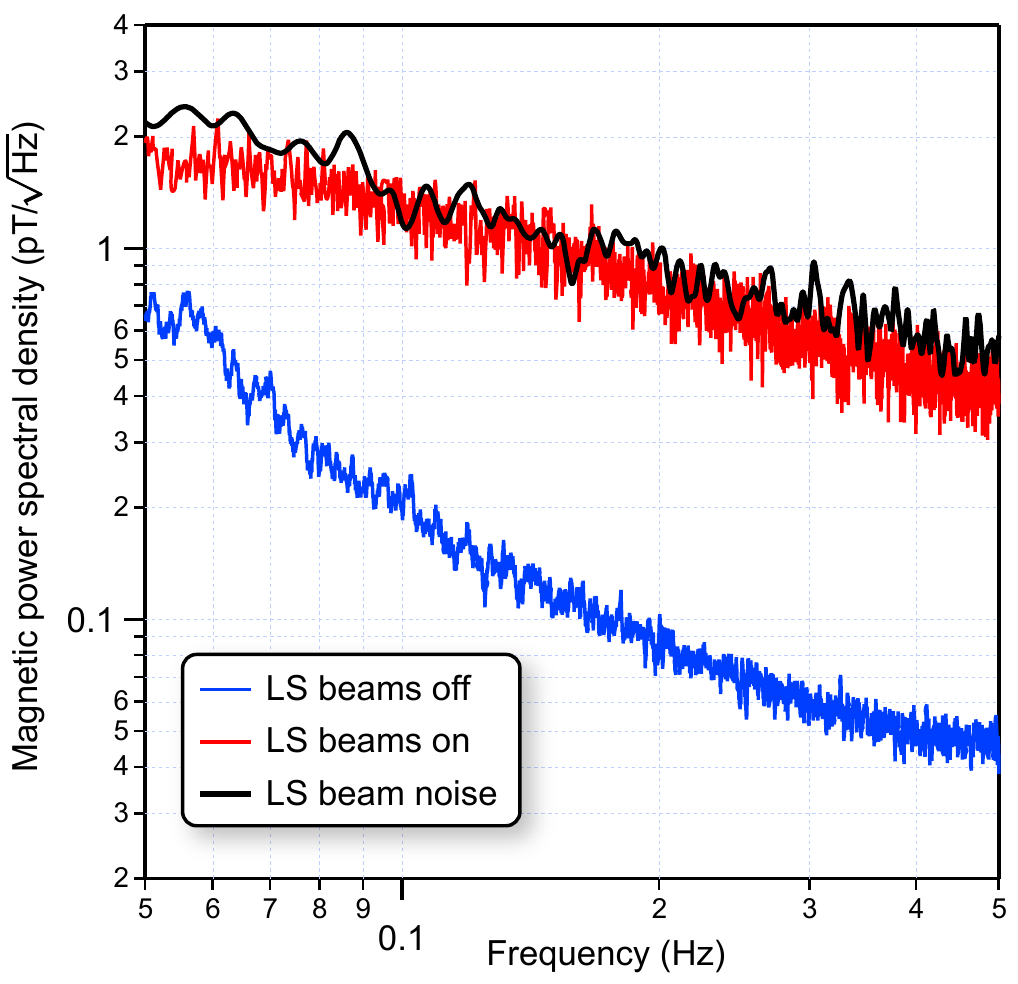}
{\caption[Measured magnetic-field PSD.]{\label{PSDplot} Power-spectral-density plot of the scalar field measurement with the $LS_z$ beams turned off (blue) and turned on at a constant power of 1 mW (red).  For these data, $\theta = 0$ and the light-shift beam power was not actively controlled.  The black trace is the predicted noise floor the scalar field measurement taken from a (separate) recording of the light-shift beam power, from which a PSD was derived and the effective magnetic field calculated using the observed light-shift coefficients $\alpha_{y}$ and $\alpha_{z}$.}}
\end{figure}

The data shown in Fig.~\ref{tanplot} were fit to Eq.~(\ref{taneq}).  The best-fit ratio $(P_y \alpha_y / P_z \alpha_z)$ was measured to be (${0.94 \pm 0.01}$) rather than unity, possibly due to slight asymmetry in the cell dimensions or systematic uncertainty of the beam powers within the cell.  With no added light-shift beams, the synchronously pumped scalar sensor has sensitivity of 48 fT/\sqrtHz for integration times of 1 second, as calculated from the power spectral density (PSD) of the measured magnetic field, shown in Fig.\ \ref{PSDplot}.  To confirm this sensitivity in the time domain, we stepped the local oscillator frequency by $\pm$0.875 mHz around $\omega_L$ and observed shifts in the lock-in \textsf{Y} output with a signal-to-noise ratio of 7.2.  Given the lock-in's equivalent noise bandwidth (ENBW) of 1.25 Hz, this corresponds to a sensitivity of 62 fT/\sqrtHz.  To assess the uncertainty in the field angle, we recorded data with the $\hat{z}$ field held constant and the $\hat{y}$ field toggled between two small values. The lock-in \textsf{Y} signal was demodulated at $\omega_y$ and $\omega_z$, and the ratio of these two responses converted to a measured magnetic-field angle according to the best-fit curve shown in Fig.~\ref{tanplot}.  The resulting plot of $\theta$ vs.\ time is shown in Fig.~\ref{StepPlot}.  The modulation of the field angle is clearly visible, and the rms noise in the ratio corresponds to 0.47 mrad/\sqrtHz precision in the measured angle of the magnetic field.  (This takes into account the measured ENBW of the software demodulation procedure.)

In the present setup, the precision of the measurement of $\theta$ is limited by apparent magnetic noise induced by fluctuations in the light-shift beam powers.  With $LS_z$ set to 1 mW without modulation and the field along $\hat{z}$, the smallest observable magnetic-field step with 1 Hz ENBW was 1.3 pT -- a factor of 21 worse than the same data recorded without the light-shift beams.  Power fluctuations in the $LS_y$ and $LS_z$ beams were recorded and converted into effective magnetic-field fluctuations according to the observed light-shift coefficients $\alpha_{y,z}$.  As shown in Fig.\ \ref{PSDplot}, the predicted magnetic noise floor matches that observed in the magnetic-field PSD.  Better control of intensity noise within the light-shift beams should allow dramatically improved scalar measurements and correspondingly better sensitivity to the field angle.  The scalar sensitivity of the magnetometer would be 12 fT/\sqrtHz if the polarimeter and amplifiers operate at the photon shot-noise limit.  By eliminating these sources of technical noise, it should be possible to reach a sensitivity of \mbox{5 $\mu$rad/\sqrtHz} in the measurement of the magnetic-field direction.   

Expanding the vector measurement to three dimensions will simply require adding another light-shift beam in the $\hat{x}$ direction.  The bandwidth of the vector measurement is presently limited by the narrow magnetic-resonance line, but this can be expanded by power-broadening the resonance with the probe beam or heating the cell to increase the Cs density and spin-exchange-broadened linewidth.  Either technique would allow more rapid measurement of the vector field components with little if any loss in sensitivity.  As discussed in the Supplemental Material, the uncertainty in the measured angle $\theta$ has no intrinsic dependence on the magnitude of the ambient field $B_0$.  Consequently, this technique should be applicable for vector magnetometry in geophysical fields with comparable precision, provided that a similar scalar sensitivity can be achieved.



In summary, we have demonstrated a method for measuring the magnitude and direction of a magnetic field through all-optical interrogation of an atomic sample.  This technique offers advantages over other methods (such as EIT vector magnetometry) because it relies on measuring changes in the magnetic-resonance frequency, rather than resonance amplitudes which can be affected by many experimental factors.  Further optimization of the apparatus will allow for a compact, magnetically inert vector magnetometer well-suited for precision physics experiments or geophysical field measurement.

We thank Douglas Beck, Michael Sturm, David Wurm, and Peter Fierlinger for their helpful input, as well as Mikhail Balabas for preparation of the antirelaxation-coated cesium cell.  We also thank Arne Wickenbrock for contributions to the measurement.  This work was funded in part by a University of California UC Discovery Proof of Concept grant (award 197073) and NASA SBIR contract NNX13CG20P, and is supported by National Science Foundation Grant No. PHY-1068875.  B. Patton is supported by DFG Priority Program SPP1491, `Precision measurements with cold and ultracold neutrons'.

\newpage

\bibliography{VectorMagBib}
\bibliographystyle{prsty}

\end{document}